\documentclass[twocolumn]{aastex63}

\usepackage{float}
\usepackage{amsmath,amssymb}
\usepackage{natbib}    
\usepackage{hyperref} 
\usepackage{graphicx} 
\usepackage{color}
\usepackage{verbatim}    
\usepackage{enumitem}
\usepackage{natbib}
\usepackage{CJK}
\usepackage{multirow}
\usepackage{lineno}

\newcommand{\proptosim}{\mathrel{\vcenter{
 \offinterlineskip\halign{\hfil$##$\cr
 \propto\cr\noalign{\kern2pt}\sim\cr\noalign{\kern-2pt}}}}}

\newcommand{\au}{\mathrm{AU}}

\newcommand{\K}{\mathrm{K}}  
\newcommand{\km}{\mathrm{km}}

\newcommand{\erg}{\mathrm{erg}}

\newcommand{\s}{\mathrm{s}}
\newcommand{\yr}{\mathrm{yr}}

\renewcommand{\d}{\mathrm{d}}

\newcommand{\p}{{\rm p}}

\newcommand{\figdir}{.}

\begin{document}
 
\title{Dynamical Effects of Colliding Outflows in Binary
  Systems}

\author[0000-0002-6540-7042]{Lile Wang}
\affil{The Kavli Institute of Astronomy and Astrophysics,
  Peking University, Beijing 100084, China; 
  lilew@pku.edu.cn}
\affil{Center for Computational Astrophysics, Flatiron
  Institute, 162 5th Ave, New York, NY 10010, USA}

\author[0000-0003-0750-3543]{Xinyu Li}
\affil{Canadian Institute for Theoretical Astrophysics, 60
  St George St, Toronto, ON, M5R 2M8, Canada}
\affil{Perimeter Institute for Theoretical Physics, 31
  Caroline Street North, Waterloo, ON, N2L 2Y5, Canada}

\begin{abstract}
  The outflow of an object traveling in a fluid can shape
  the fluid morphology by forming a forward bow shock which
  accelerates the object via gravitational feedback. This
  dynamical effect, namely ``dynamical anti-friction'', has
  been studied in idealized infinite uniform media, which
  suffers from the convergence problem due to the long-range
  nature of gravitation. In this work, we conduct global 3D
  hydrodynamic simulations to study this effect in the
  scenario of a binary system, where the collision of
  outflows from both stars creates a suitable
  configuration. We demonstrate with simulations that a
  dense and slow outflow can give rise to a positive torque
  on the binary and lead to the expansion of the orbit. As
  an application, we show that binaries consisting of an AGB
  star and an outflowing pulsar can experience $\sim 10~\%$
  orbit expansion during the AGB stage, in addition to the
  contribution from mass-loss. We also prove that the
  gravitational force drops as $O(r^{-3})$ from the center
  of mass in the binary scenarios, which guarantees a quick
  converge of the overall effect.
\end{abstract}


\section{Introduction} \label{sec:intro}

Winds and outflows are launched from stellar objects through
various mechanisms. For high-mass stars, winds are usually
radiatively driven through line absorption
\citep{1999isw..book.....L} with speed up to
$\gtrsim 10^3$~km/s and mass loss rate larger than
$10^{-5}$~$M_\odot$/yr. For low-mass stars, the mechanism of
driving wind are likely analogous to the solar wind, which
is accelerated by the thermal pressure gradient
\citep{1958ApJ...128..664P} with a much lower mass-loss rate
($10^{-14}$~$M-\odot$/yr for the solar wind). As a low or
intermediate-mass star evolves into the asymptotic giant
branch (AGB) phase, slow ($\sim10$~km/s) and dense
($\dot{M}\sim 10^{-8}-10^{-4}$~$M_\odot$/yr) are launched
from the star as the fast wind originating from the stellar
remnants interacts with the inner layer of the expanding
stellar envelope \citep {2004agbs.book.....H,
  2008A&A...487..645R}. Besides, strong outflows can be
launched from the compact objects either by the strong
magnetic fields in the form of pulsar winds or by
super-Eddington accretion in the case of SS433
\citep{2004ASPRv..12....1F}.

In an isolated binary system where a secondary star is
orbiting around the primary star, the secondary star exerts
a gravitational drag on the wind materials from the primary
and accretes a fraction of it. This accretion process of
wind materials can occur in high-mass X-ray binaries (see
\citealt{2015ARep...59..645S} for a review on wind
accretion) and is described by the standard
Bondi-Hoyle-Lyttleton (BHL) accretion
\citep{1939PCPS...35..405H, 1944MNRAS.104..273B,
  2004NewAR..48..843E}. Incoming materials with impact
parameters inside a cylinder of the Bondi
radius \begin{equation} R_B =
  \frac{2GM}{c_s^2+V_*^2} \end{equation} are accreted onto
the star, while materials with greater impact parameters
accumulate behind the star, forming an over-dense wake which
acts gravitationally to decelerate the star, knows as the
dynamical friction \citep {1943ApJ....97..255C,
  1999ApJ...513..252O}. Here, $G$ is the Newtonian
gravitational constant, $M$ and $V_*$ are the mass and
velocity of the star, respectively, and $c_s$ is the sound
speed of the ambient materials. As the secondary star moves
around the host star, the over-dense wake is trailing behind
the orbit and is expected to dampen the orbital motion and
shrink the orbital separation of the two stars. Previous
analytical and numerical study
\citep{2010MNRAS.402.1758S,2019ApJ...884...22A} of binary
accretion over a dense gaseous medium confirmed the orbital
decay due to dynamical friction.

The picture of dynamical friction needs to be reconsidered
when the secondary star also has outflows. The collision of
the strong outflow from the secondary with the incoming
materials forms a bow shock around of the secondary
\citep{1996ApJ...459L..31W,2005AdSpR..35.1116G}. Incoming
materials will stream along the shock front away from the
secondary, and the accretion is shut off, leaving a
low-density bubble behind the
secondary. \citet{2020MNRAS.492.2755G} calculated the
gravitational force from the bow shock and found that the
net force can accelerate the star given the outflow is
strong enough. This phenomenon is proved by numerical
simulations \citep{2020MNRAS.494.2327L} for a star moving in
an idealized uniform gas. Similar effects in the planetary
discs where thermal luminosities injected from the hot
planet lead to a net force opposite to the dynamical
friction \citep{2017MNRAS.465.3175M, 2017MNRAS.472.4204M,
  2017MNRAS.469..206E, 2020ApJ...902...50H}.

However, previous numerical studies on dynamical friction or
anti-friction are usually set up as a local ``wind-tunnel''
simulation \citep[e.g.][] {2017ApJ...838...56M,
  2020MNRAS.494.2327L, 2020ApJ...897..130D} with an
idealized infinite background gas of uniform density or with
density gradient. This setup suffers from the convergence
problem caused by the long-range nature of gravitation. No
matter how far away, matters behind the star will be
gravitationally dragged by the star and decelerate the
star. The deceleration increases logarithmically with the
size of the computational box. Therefore results with
uniform gas cannot draw realistic conclusions on the orbital
evolution of the binary. Global simulations with realistic
gas distribution are required and have been applied to study
dynamical friction \citep{2021arXiv210709675S}.

This paper performs 3D global realistic simulations of a
binary system with outflows launched from both stars. We
focus on how the interactions between the outflows shape the
orbital evolution of the binary systems. Unless specified,
we present our results with all physical quantities in the
code unit with mass $m_c$, length $\ell_c$, and time $t_c$.

This paper is structured as follows. In \S\ref{sec:method}
we briefly introduce the numerical methods, the general
setups of problems, and some key diagnostic quantities for
post-simulation analyses. \S\ref{sec:model-equimass}
discusses an idealized binary that consists of two nearly
identical stars and explores the impact from different
physical parameters using a few series of extra models. We
apply our calculation to a specific model of an AGB-pulsar
binary in \S\ref{sec:agbp}. Discussion and conclusion are
presented in \S\ref{sec:discussions} where we prove that we
obtain robust results which are convergent and independent
of the computational box size.

\section{Methods}
\label{sec:method}

\subsection{Basic Geometry and Configurations}
\label{sec:method-setup}

\begin{figure}
  \centering
  \hspace*{-0.1in} 
  \includegraphics[width=3.2in, keepaspectratio]
  {\figdir/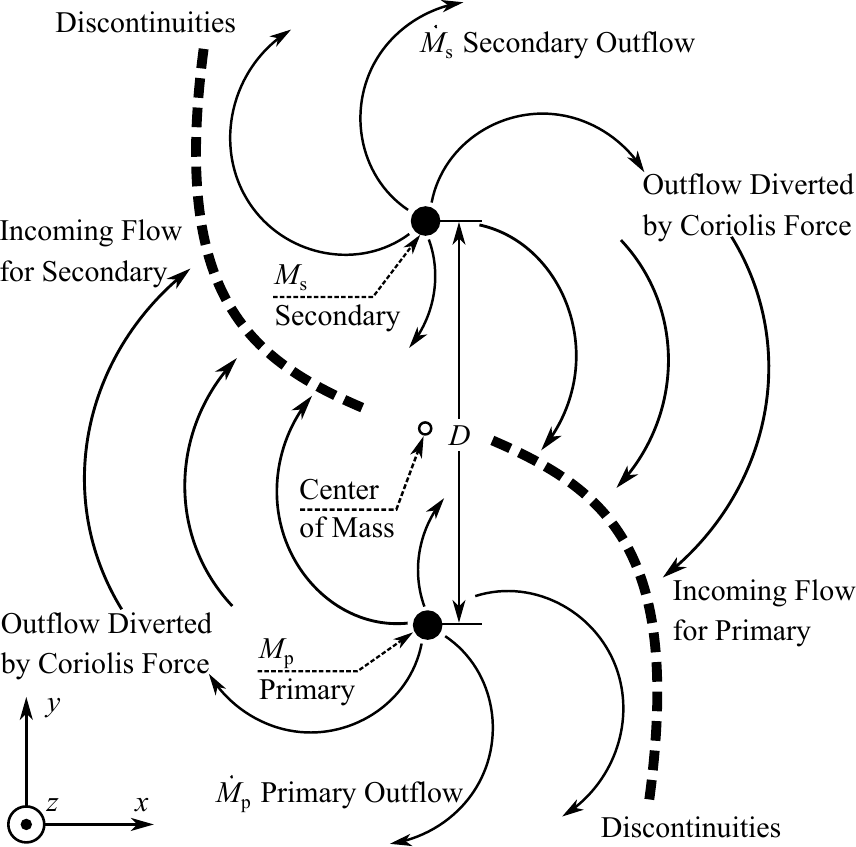}
  \caption{ Simulation setups in the co-rotating frame. The
    two stars with mass $M_{\rm p,s}$ are fixed on the
    $y$-axis and separated by distance $D$.  Outflows are
    launched from the two stars with mass-loss rates
    $\dot{M}_{\rm p,s}$. Details of the outflow colliding
    regions are omitted in this figure.  We simply annotate
    ``Discontinuities'' generically for the relevant
    phenomena.}
  \label{fig:setup-schematic}
\end{figure}

We simulate the evolution of a binary system consisting of a
primary star and a secondary star on a Cartesian grid in
their {\it co-rotating frame}. The setup is illustrated by
Figure~\ref{fig:setup-schematic}. The subscripts ``p'' and
``s'' denote the quantities associated with the primary and
the secondary star throughout this paper, respectively.

The two stars are located along the $y$-axis, separated by
distance $D$ and orbit each other on a circular orbit. For
convenience, we denote the star on the negative half of the
$y$-axis as the primary, and the one on the positive half as
the secondary with mass $M_{\p, \s}$ respectively. The
$y$-coordinates of the stars are set as
$y_\p=-D M_\s/(M_\s + M_\p)$ and $y_\s=D M_\p/(M_\s+M_\p)$,
so that the center of mass is fixed at the origin.

Rectangular boxes define the simulation domains in this
paper, in which Cartesian grids are used. The orbital
angular velocity vector is parallel to the $z$-axis, so that
the orbit stays in the $z=0$ plane. Due to the system
geometry, we assume reflection geometry over the $z=0$ plane
and adopt a reflecting boundary condition at that plane;
outflowing boundary conditions are adopted elsewhere. The
simulation domain size is defined by box size $L_{\rm box}$:
the domain spans a sub-space
$\{x, y, z\}/(L_{\rm box}/2)\in [-1, 1]\otimes [-1,
1]\otimes [0, 1]$. We denote the resolution of the unrefined
base grid with $N_0$ as
$N_x\times N_y\times N_z = N_0\times N_0\times (N_0/2)$,
which yields unitary aspect ratios.  We set up multiple
levels of static mesh refinement near the source regions to
resolve each source radius by at least three cells. All
simulations are run till steady or quasi-steady states are
reached.

\subsection{Numerical Hydrodynamics}
\label{sec:method-numerical}

We use a newly developed, GPU-optimized code \verb|Kratos|
(L. Wang, in prep.) to numerically evolve the hydrodynamic
equations for the dynamics of the outflowing gas,
\begin{equation}
  \label{eq:hydrodyn}
  \begin{split}
    \dfrac{\partial \rho}{\partial t} + \nabla \cdot (\rho
    \mathbf{v}) & = 0\ ,\\
    \dfrac{\partial (\rho \mathbf{v})}{\partial t}
    + \nabla \cdot (\rho \mathbf{v} \mathbf{v} +
    p\mathbf{I}) & = -\rho \nabla\Phi ,\\
    \dfrac{\partial \varepsilon}{\partial t}
    + \nabla \cdot [\mathbf{v}(\varepsilon + p) ] & = -\rho
    \mathbf{v}\cdot \nabla\Phi\ .
  \end{split}
\end{equation}
Here, $\rho$, $\mathbf{v}$, $p$ and $\epsilon$ are the mass
density, velocity, gas pressure, and total energy density,
respectively, $\Phi$ is the gravitational potential, and
$\mathbf{I}$ is the identity tensor. In the \verb|Kratos|
framework, hydrodynamic solvers are employed by the
higher-order Godunov method, involving piecewise-linear
method (PLM) reconstruction with minmod slope limiter, HLLC
Riemann solver, and second-order Runge-Kutta time integrator
for second-order accuracy in both space and time \citep[see
e.g.][]{Toro2011}.

Eqs.~\eqref{eq:hydrodyn} are integrated in the
co-rotating frame with the binary angular frequency
$\Omega = [G(M_\p+M_\s)/D^3]^{1/2}$.  The centrifugal
acceleration $\mathbf{a}_{\rm cf}$ and Coriolis acceleration
$\mathbf{a}_{\rm co}$ in this rotating frame read,
\begin{equation}
  \label{eq:centfug-coriolis}
  \mathbf{a}_{\rm cf} = \mathbf{\Omega} \times(\mathbf{x}
  \times \mathbf{\Omega})\ ,\quad
  \mathbf{a}_{\rm co} = -2\mathbf{\Omega}\times\mathbf{v}\ ,
\end{equation}
which are fed as the additional source terms into
eq.~\eqref{eq:hydrodyn}.  Here $\mathbf{\Omega}$ is the
angular velocity vector, and we denote $P$ to be the orbital
period. Note that these two terms are specific to Cartesian
meshes; their effects can be included in the simulations in
spherical-polar and cylindrical coordinates naturally
without actually referring to the source terms \citep[see
also][]{1998A&A...338L..37K}.

The two stars source the gravitational potential for
eq.~\eqref{eq:hydrodyn} and we neglect the self-gravity of
the gas. Both stars also serve as ``source particles''
\citep[see e.g.][]{2017MNRAS.465.1316M} to launch isotopic
outflows. In the adjacency of each star, we set up a
spherical ``source region'' with radius $r_{\rm src}$, in
which an initial radial velocity $v_{\rm src}$ is set as a
model parameter. The density is set to produce the desired
mass-loss rate $\dot{M}=4\pi r^2\rho v_{\rm src}$. The gas
pressure is set based on a chosen temperature parameter
$p/\rho$. The physical quantities are fixed during the
hydrodynamic evolution, producing constant outflows from the
source regions.

We adopt the ideal equation of state for the gas with a
fixed ratio of heat capacity $\gamma$. It is well-known that
an adiabatic hydrodynamic system with a central point source
of gravity cannot sustain steady supersonic outflows if
$\gamma \geq 3/2$
\citep[e.g.][]{1958ApJ...128..664P}. Despite the actual
states of materials, such outflows could remain
$\gamma < 3/2$ due to various heating mechanisms in
reality. For simplicity, we adopt a specific $\gamma$ for
each problem within this paper.

\subsection{Analyses Schemes}
\label{sec:method-ana} 

After reaching supersonic regimes, perturbations of fluids
can only feedback their impact onto the objects by
gravity. We calculate the gravitational acceleration of each
star from the gas component by summing over every piece of
fluid mass \begin{equation} \label{eq:grav-acc}
  \mathbf{g}_{\p,\s} = G \int_V \d V \dfrac{\rho\,
    \mathbf{x}_{\p,\s}} {|\mathbf{x}_{\p,\s}|^3}\
  , \end{equation} where $\mathbf{x}_{\p,\s}$ are the
direction vectors pointing from the primary or the secondary
to the gas element, and the integral runs over the whole
simulation domain. Since our simulation box only contains
the upper half-space, we make use of the reflection symmetry
and add the contribution of $\mathbf{g}$ from the lower
half-space.

With the fundamental configurations in
Figure~\ref{fig:setup-schematic}, the secular orbital
evolution is only affected by the tangential component of
the gravity, $g_\tau$. Constant perturbations on the radial
force does not affect any orbital parameters, and the
reflection symmetry of the system guarantees zero net force
parallel to the axis of angular velocity. Summing up the
torques on the two stars, we obtain the relative rate
$\Gamma$ at which the orbital scale $D$ varies,
\begin{equation}
  \label{eq:dlna-dt}
  \Gamma \equiv \dfrac{\dot{D}}{D} = \dfrac
  {2 ( T_{\rm p} + T_{\rm s})}{ L_{\rm p} + L_{\rm s}}
  = \dfrac{2D}{ \Lambda}  (g_{\rm \tau; p} + g_{\rm \tau; s})\ .
\end{equation}
Here, $T_{\p, \s} = \mu g_{\tau; \p,\s} D$ are the torques
on the primary or secondary, $L_{\p, \s}$ are the angular
momenta with respect to the center of mass,
$\mu \equiv M_\p M_\s / (M_\p + M_\s)$ is the reduced mass,
and $\Lambda = [G D (M_\p + M_\s)]^{1/2}$ is the reduced
specific angular momentum which satisfies
$L_\p + L_\s = \mu \Lambda$.  Eqs.~\eqref{eq:grav-acc} and
\eqref{eq:dlna-dt} evaluate the key effect concerned in this
paper.

To quantify the contribution to the orbit expansion from gas
element at different locations, we define $\tau$ to be the
``torque density'' as,
\begin{equation}
  \label{eq:tau-torque}
  \tau 
  = G D\rho\mu \left(\dfrac{\delta x_\p}{r_\p^3} -
    \dfrac{\delta x_\s}{r_\s^3} \right)\ ,
\end{equation}
where $\delta x_{\p, \s}$ and $r_{\p, \s}$ are the
$x$-direction signed displacement and the distance, measured
from the primary and the secondary, respectively. The sign
is chosen in such a way that a positive $\tau$ leads to
positive contribution to $\Gamma$. It is also
straightforward to verify that
$\int_V\tau \d V = T_\p + T_\s$.

\section{
  Near-Identical Binaries}
\label{sec:model-equimass}

This section discusses a type of model binaries whose both
stars have similar conditions.

\begin{figure*}
  \centering
  \hspace*{-0.1in} 
  \includegraphics[width=6.5in, keepaspectratio]
  {\figdir/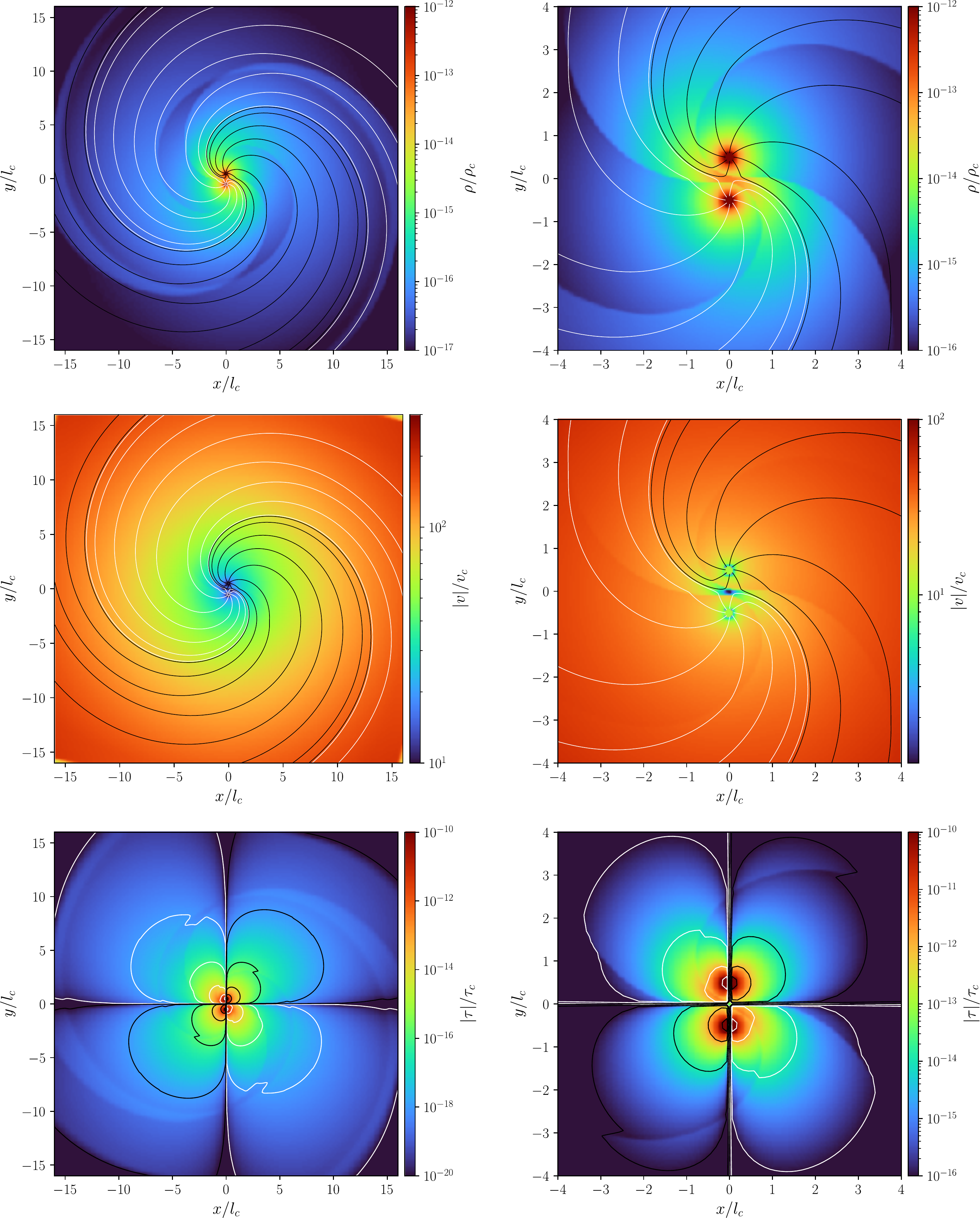}
  \caption{Key hydrodynamic quantities in the orbital plane
    for the $t = 5.5~P$ snapshot of Model EM
    (\S\ref{sec:equimass-fiducial}), including mass density
    $\rho$ (upper row), the magnitude of co-rotating frame
    velocity $|v|$ (middle row), and the magnitude of
    ``torque density'' $|\tau|$ (lower row; see also
    \S\ref{sec:method-ana}, eq.~\ref{eq:tau-torque}), all
    normalized by the code units (subscript ``$c$'';
    $\rho_c\equiv m_c \ell_c^{-3}$,
    $v_c\equiv \ell_ct_c^{-1}$, and
    $\tau_c\equiv m_c \ell_c^{-1}t_c^{-2} = p_c$).  The left
    column shows the full cross-section of the domain
    ($32~\ell_c\times 32~\ell_c$), and the right column
    zooms in to the $8~\ell_c\times 8~\ell_c$ near the
    origin. Note that the zoom-in plots may have {\it
      different} dynamical ranges. Streamlines are overlaid
    in each panel except the bottom; black lines are those
    initiating from the secondary, and white lines from the
    primary. The contours in the bottom row indicate the
    magnitudes (on
    $\lg|\tau/\tau_c| \in\{-20, -18, -16, -12, -10 \}$ for
    the left one, and $\{ -18, -16, -12, -10 \}$ on the
    right) and signs (white for positive; black for
    negative) of the torque denstiy. The outermost black
    contours are hardly visible.}
  \label{fig:equimass-fiducial}
\end{figure*}

\subsection{Fiducial Model Setup: Identical Stars}
\label{sec:equimass-fiducial}

We start with a fiducial model (also Model EM) that contains
two identical stars with the same outflow conditions. These
two stars with $M_\p = M_\s = m_c$ are separated by
$D = \ell_c$, and orbit each other on a circular orbit with
period $P=2\pi[G(M_\p + M_\s)/D^3]^{-1/2} = 0.707~\yr$.  The
outflows of both star are characterized by the mass-loss
rate $\dot{M}_0 \equiv 4\pi\times 10^{-13}~m_c/t_c$, initial
pressure-to-density ratio $p/\rho = 10^2~(\ell_c/t_c)^2$,
and the initial velocity
$v_{\rm src} = v_0\equiv 11~\ell_c/t_c$ within the source
region $r_{\rm src} = 0.1~\ell_c$. We set $\gamma = 1.01$ to
emulate a near-isothermal condition for the outflowing gas.
According to the Parker isothermal wind solution, the
outflow should reach $v\sim 33~\ell_c/t_c$ when it reaches
the approximates of the companion. We use the box size
$L_{\rm box} = 32~\ell_c$ and the base resolution of $256$
zones per $L_{\rm box}$. Two levels of static mesh
refinement are set near the centers of both stars, so that
the source region radius $r_{\rm src}$ is resolved by no
less than $3$ zones.

A convergence test (marked as Model EM-C) is conducted, in
which the base resolution is $512$ cells per $L_{\rm box}$,
and the number of levels for mesh refinement is kept as 2,
to double the resolution everywhere. We also perform another
test run (denoted by Model EM-SO) in which the outflow of
the secondary is turned off totally. These simulations are
run for $t\gtrsim 5.5~P$ to guarantee the convergence to the
(quasi-)steady states. The fiducial run takes $\sim 12$
minutes of wallclock time on a workstation with two NVIDIA
RTX 3090 GPUs.

\begin{figure}
  \centering
  \hspace*{-0.1in} 
  \includegraphics[width=3.4in, keepaspectratio]
  {\figdir/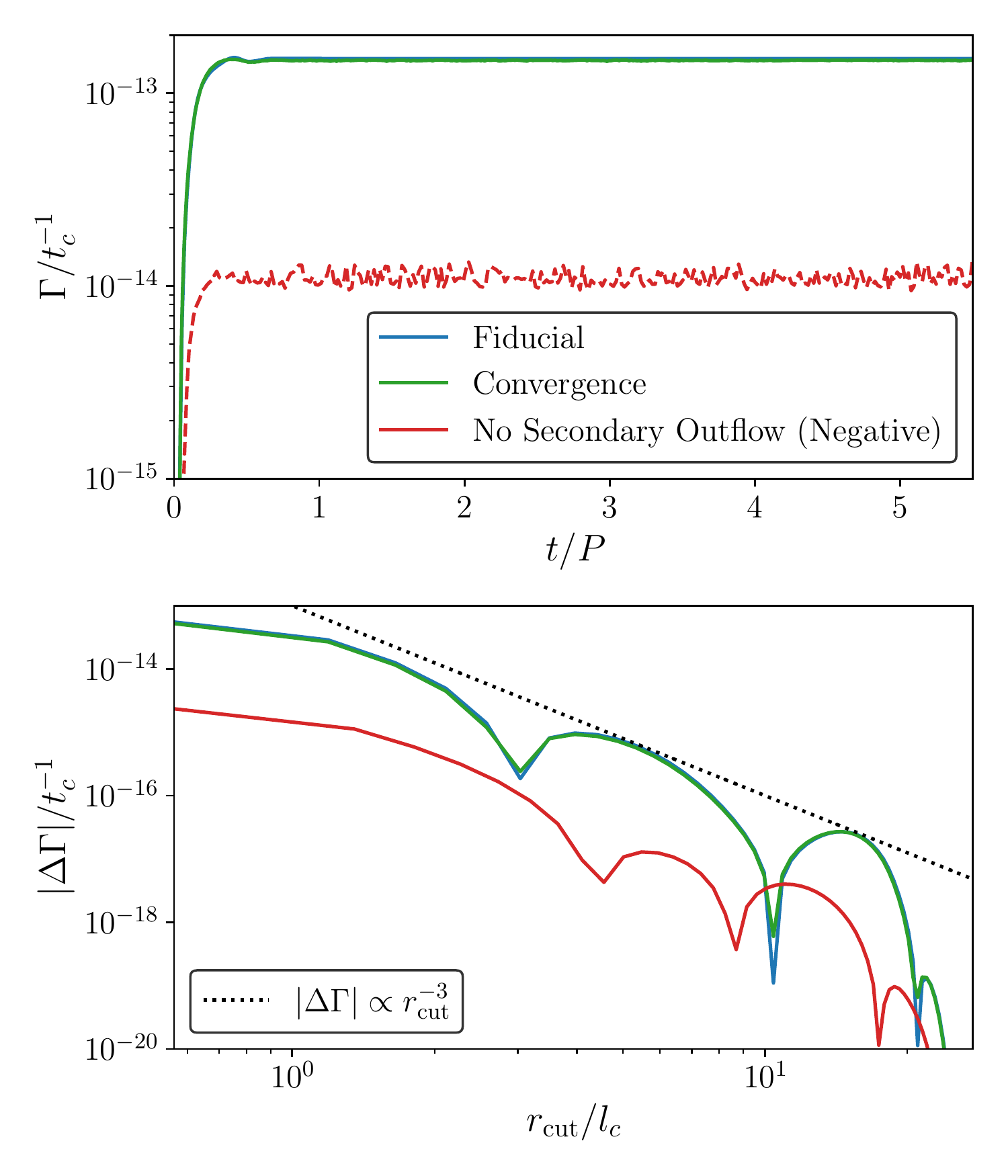} 
  \caption{{\bf Upper panel}: relative rate of orbit
    expansion $\Gamma \equiv \dot{D}/D$ for the fiducial
    model (Model EM) and its associated tests (distinguished
    by colors) in \S\ref{sec:equimass-fiducial-result}; {\bf
      negative} numbers are in {\bf dashed lines}. For
    comparison, the double-resolution convergence test, and
    the test case with the secondary outflow turned {\it
      off} are also presented in different colors. Note that
    the fiducial model and the convergence test overlap
    almost perfectly. {\bf Lower panel}: Residual of orbit
    expansion rate as a function of cutoff radius
    $r_{\rm cut}$ (measured from the center of mass),
    compared to the full-domain value (see also 
    \S\ref{sec:equimass-fiducial-result}). A dotted line is
    also plotted to illustrate the $\sim r_{\rm cut}^{-3}$
    scaling. }
  \label{fig:dlns-equimass-fiducial}
\end{figure}

\subsection{Fiducial Model Results}
\label{sec:equimass-fiducial-result}

We present in Figure~\ref{fig:equimass-fiducial} the key
hydrodynamic profiles on the orbital plane at $t = 5.5~P$
for the fiducial model. The collision of identical outflows
creates spiral patterns due to orbital motion. As a result,
the qualitative features of fluid configuration required by
the dynamical anti-friction effect--a relatively over-dense
region between the bow shock and terminal shock, an
under-dense region behind the star--indeed yields a torque
that increases the angular momentum. We note that the
contact discontinuity vanishes by decaying into a continuous
contact surface. The two outflows are identical, and no
discontinuities appear at the normal collision
surfaces. Such continuity remains on both sides when
streamlines develop near the contact surface, as the
streamlines from both stars have tangential contact when
they meet.

The fiducial model orbit expansion rate is shown in the
upper panel of Figure~\ref{fig:dlns-equimass-fiducial}, in
which we find the rate of orbit increase is $\Gamma\simeq
1.51\times 10^{-13}~t_c^{-1}$. As we can observe from the
bottom-right panel, the positive torque (indicated by the
``lobes'' enclosed by white lines) overcomes the
``drawbacks'' (enclosed by black lines) mainly by the
materials near the discontinuity. Model EM-C is only
different by $\sim 2~\%$, indicating a good convergence in
terms of spatial resolution. Model EM-SO, on the other hand,
produces $\Gamma \simeq -1.1\times
10^{-14}~t_c^{-1}<0$. This result confirms that the outflows
from the secondary are essential for dynamical
anti-friction. The secondary star will migrate inward under
the normal dynamical friction mechanism without the
secondary outflow.

The lower panel of Figure~\ref{fig:equimass-fiducial}
examines the spatial convergence of orbit expansion rate by
showing the residual of
$\Delta\Gamma\equiv|\Gamma-\tilde{\Gamma}(r_{\rm cut})|$ for
different ``cutoff'' radius $r_{\rm cut}$. Here
$\tilde{\Gamma}(r_{\rm cut})$ is evaluated by
eq.~\eqref{eq:dlna-dt}, but with the integration zones
limited to regions within the distance $r < r_{\rm cut}$
from the center of mass. The $|\Delta\Gamma|$ profile is
enveloped by the $\sim r_{\rm cut}^{-3}$ power-law, shown in
the dotted line. This scaling clearly shows that our
simulations will produce a finite, converging orbit
expansion rate as we increase the computational box size;
the result does not suffer from the divergence caused by
idealized uniform background ambient. We will revisit this
in \S\ref{sec:discussion-converge} to prove the scaling law
and verify the actual convergence of our calculations.

The rate of orbit expansion in physical units is relatively
slow. If we specify the code units by $t_c = 1~\yr$,
$\ell_c = 1~\au$, and $m_c = 1~M_\odot$, this rate leads to
a $\Gamma \simeq 0.15~\%$ expansion during the $10^{10}~\yr$
(the main-sequence lifetime of a typical solar-like
star). The ``cost'' that each star pays for this expansion
during this period is $\sim 10^{-2}~M_\odot$, which would
result in another $\Gamma \sim 1~\%$ orbit expansion. In
short, this fiducial model exhibits a case in which
dynamical anti-friction enhances the orbital expansion from
mass-loss by $\sim 15\%$.

\subsection{Exploring the Parameter Space}
\label{sec:equimass-var-models}

We conduct a series of simulations to study how the
dynamical anti-friction effects depend on the key
hydrodynamic and orbital parameters.  Only one parameter is
modiefied for each model, and the values of the modified
parameters with the corresponding results are presented in
Table~\ref{table:equimass-var-model}.

In the ideal scenario that an outflowing star plunge into a
uniform gas, the acceleration of dynamical anti-friction for
strong outflow asymptotes to
\citep[see][]{2020MNRAS.492.2755G}
\begin{equation}
  \label{eq:acc-ideal}
  a_{\rm AF}\sim 8.18 ~ G\rho_0 R_0.
\end{equation}
Here $\rho_0$ is the ambient gas density, and  $R_0$ is the
distance of the contact discontinuity ahead of the star,
where the momentum flux from the outflow balances the
incoming gas.  In the scenario of the binary, we expect the
anti-friction to depend on two main factors similar to the
case of uniform gas: (1) the ambient gas density ahead of
the star where discontinuities are formed, and (2) the
balance of the momentum flux between the outflow and the
ambient gas, which determines how far away the
discontinuities reside.

Models in Series 1 vary the mass-loss rate $\dot{M}$ with
invariant $v_{\rm src}$ for both stars. Note that in steady
states the scaling $\rho(r)=\dot{M}/4\pi r^2 v(r)$
holds. Essentially, the outflow density changes
proportionally as well as the momentum flux
$\dot{M}v(r)$. However, the momentum flux ratio for two
outflows is kept the same, and the fluid configuration
should remain similar. Therefore, the force and torque from
the anti-friction are expected to scale linearly with
$\dot{M}$, as confirmed by the results in
Table~\ref{table:equimass-var-model}.

Models in Series 2 keep the same $\dot{M}$ and vary $v_{\rm
  src}$ for the two stars. The outflow density decreases
with increasing $v_{\rm src}$. We find that the orbit
expansion rate $\Gamma$ drops with an increasing $v_{\rm
  src}$ as expected. There is extra complexity in this
case. The increasing outflow velocity makes the acceleration
vector more radial rather than tangential, as the tangential
component of the ``headwind'' should always be Keplerian.

Series 3 varies the adiabatic index $\gamma$. The expansion
rate is found to increase with $\gamma$. Greater $\gamma$
gives rise to more substantial deceleration of the outflows
as they expand outward; therefore, smaller radial velocities
but greater densities are met when the outflows reach the
interaction region.

In Series 4, we vary the orbital separation $D$ and find
that $\Gamma$ is anti-correlated with $D$. This is once
again the outcome of a greater tangential velocity component
as well as a higher gas density at smaller $D$.
The actual scaling shown by Series 4 together with the
fiducial model is steeper than linear, because of the
complexities brought by the binary motion. By feeding in the
physical units, the expansion rate increases super-linearly
to $\sim 0.5~\%$ per $10^{10}~\yr$ for the Model EM-DN with
$D = 0.5~\au$, which becomes comparable to the contribution
by the mass-loss itself.

Series 5 breaks the assumption of equal mass for the binary
and varies $M_{\s}$, whose results do not exhibit
significant difference compared to the fiducial model. The
outflows become supersonic when they reach the dynamically
important zones, and the gravity from the secondary is not
important. Therefore the overall torque is not sensitive to
the stellar mass.

\renewcommand{\arraystretch}{1.0}
\begin{deluxetable}{llcc}
  \tablecolumns{4} 
  \tabletypesize{\scriptsize}
  \tablewidth{240pt}
  \tablecaption{Various models for nearly equal
    binaries. \label{table:equimass-var-model} }
  \tablehead{
    \colhead{Series} &
    \colhead{Model} &
    \colhead{Description$^*$} &
    \colhead{$\Gamma$}
    \vspace*{-0.1in} 
    \\
    \colhead{} &\colhead{} &\colhead{} &
    \colhead{$(10^{-14}~t_c^{-1})$} 
  }
  \startdata
  0 & EM & Fiducial & 15.1 \\
  & EM-C  & Convergence Test & 14.8 \\
  & EM-SO & No Secondary Outflow & $-1.1$ \\    
  \hline
  1 & EM-OH
  & $\dot{M}_{\p,\s}=10\times \dot{M}_{0}$ & 152 \\
  & EM-OL
  & $\dot{M}_{\p,\s}=0.1\times \dot{M}_{0}$ & 1.51 \\
  \hline
  2 & EM-VH & $v_{\rm src} = 3v_0$ & 10.8 \\
  & EM-VL  & $v_{\rm src} = 0.3v_0$ & 25.6 \\
  \hline
  3 & EM-G1 & $\gamma = 1.15$ & 25.5 \\
  & EM-G2 & $\gamma = 1.4$ & 67.9 \\
  \hline
  4 & EM-DF & $D = 2\ell_c$ & 5.2 \\
  & EM-DN & $D = 0.5\ell_c$ & 48.9 \\
  \hline
  5 & VM-1 & $M_{\s}=m_c/2$   & 14.8 \\
  & VM-2 & $M_{\s}=m_c/4$ & 14.0 \\  
  \enddata
  \tablecomments{*: Describes the only parameter that it
    differes from the fiducial model.  }
\end{deluxetable}

\begin{figure*}
  \centering
  \hspace*{-0.1in} 
  \includegraphics[width=6.5in, keepaspectratio]
  {\figdir/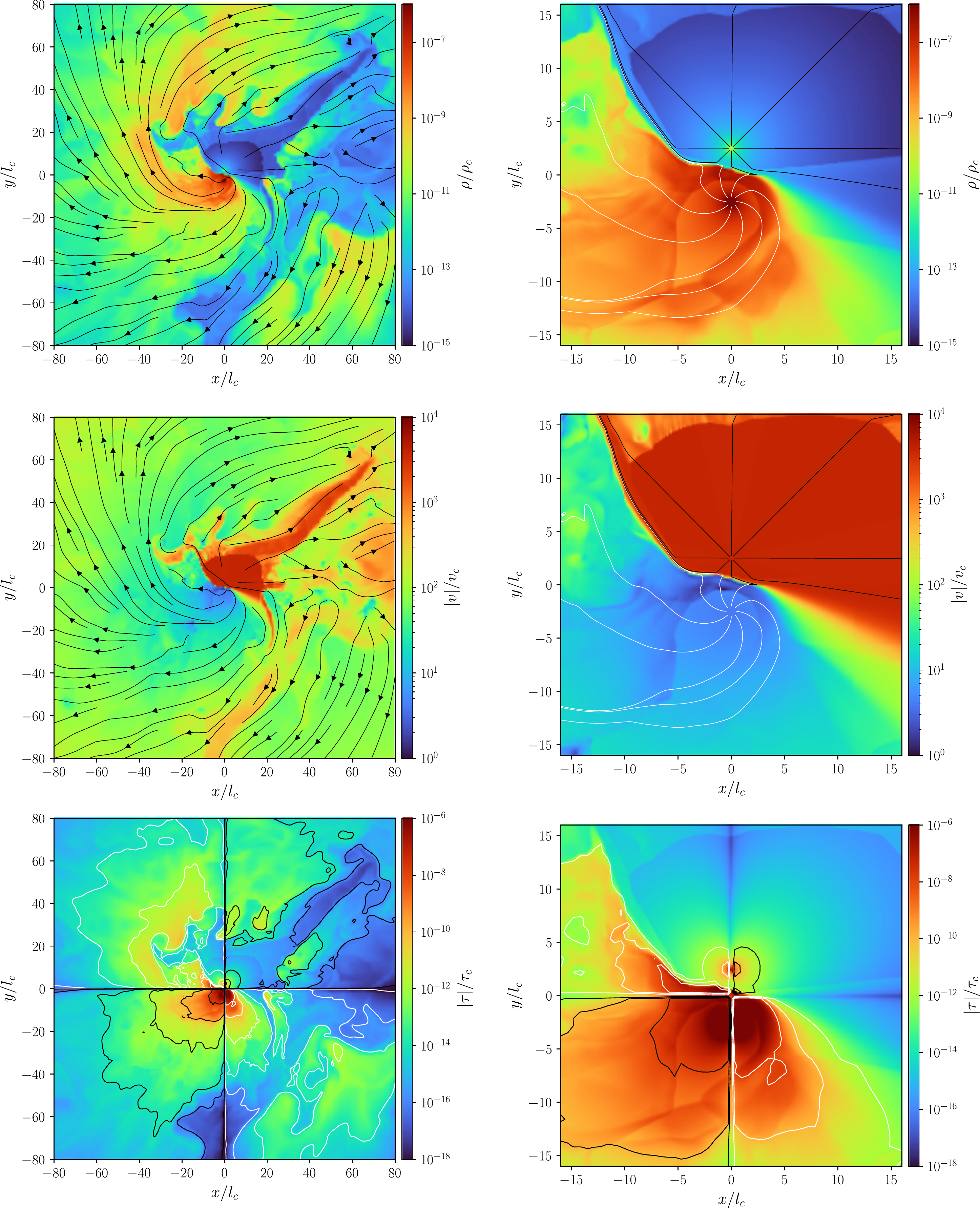}
  \caption{Similar to Figure~\ref{fig:equimass-fiducial} but
    for the fiducial AGB-pulsar binary model (Model A0) at
    the $t =7~P$ snapshot (orbital period
    $P\simeq 7.91~t_c$). Note that, because this model
    exhibits strong shear instabilities, turbulences and
    mixtures on large scales, we over-plot the streamlines
    in the left column in a different way. On the bottom
    row, the contours on the left are laid on
    $\lg|\tau/\tau_c| \in \{ -18, -16, -12, -10,-8 \}$,
    while those on the right indicate
    $\lg|\tau/\tau_c| \in \{ -12, -10,-8 \}$. }
  \label{fig:agb-pulsar-fiducial}
\end{figure*}

\begin{figure}
  \centering
  \hspace*{-0.1in} 
  \includegraphics[width=3.4in, keepaspectratio]
  {\figdir/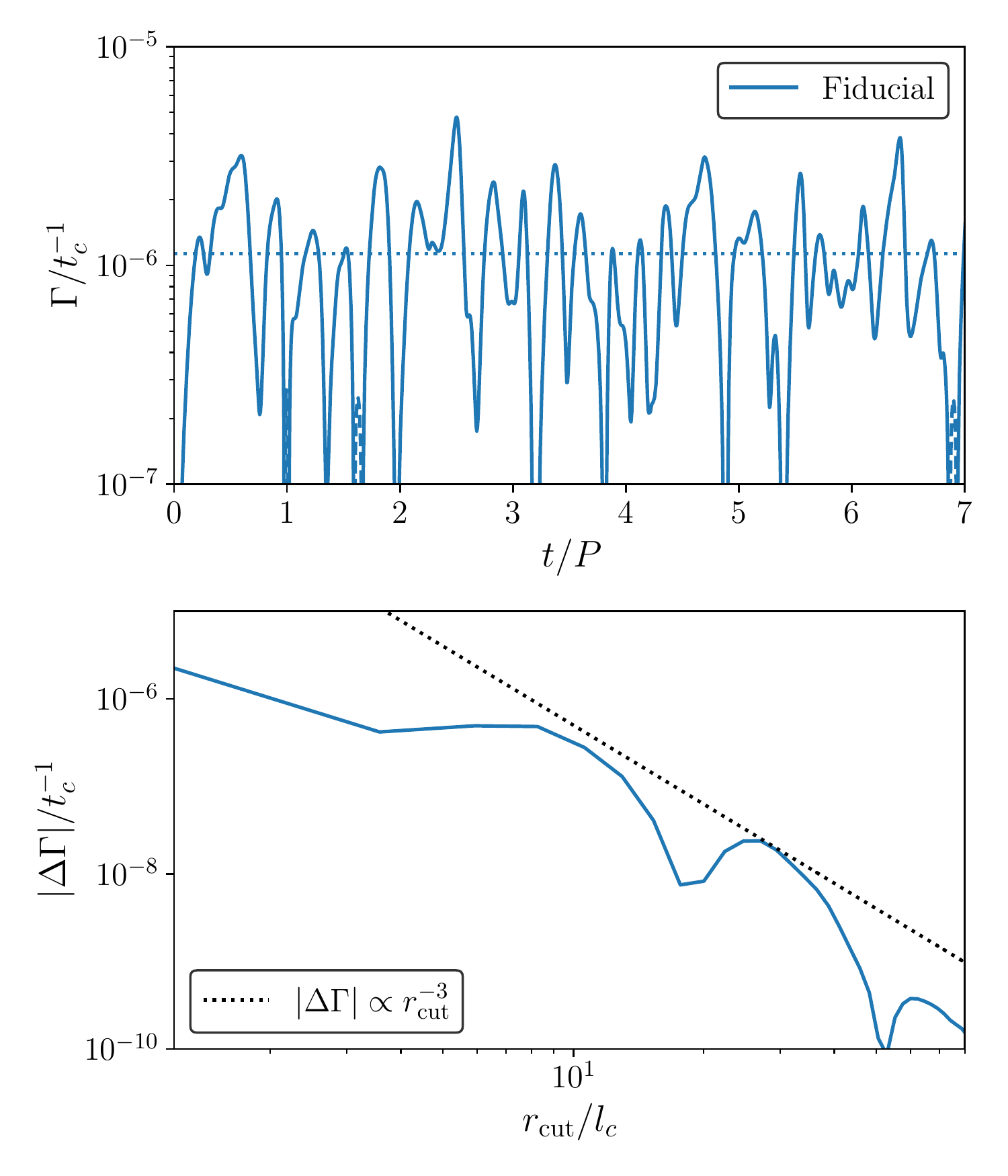}
  \caption{Similar to
    Figure~\ref{fig:dlns-equimass-fiducial} but for the
    fiducial model of AGB-pulsar binary (Model A0). A
    horizontal dotted line is added to the upper panel to
    illustrate the time-average of the orbit expansion
    rate. }
  \label{fig:dlns-agbp-fiducial}
\end{figure}

\section{AGB-Pulsar Binaries} 
\label{sec:agbp}

As an application, this section discusses a specific
situation of broad interest: an equal-mass binary consists
of a wind-blowing pulsar and an AGB star that blows a slower
but denser wind. We refer to the AGB as the primary and the
pulsar as the secondary for convenience. Previous studies of
AGB-related binaries, such as \citet{2018MNRAS.473..747C},
often consider the cases that the companion is accreting and
thus {\it loses} angular momenta by the interactions with
the AGB wind. In this section, we revisit this problem with
pulsar companions, whose outflows are usually strong enough
to create a hydrodynamic configuration that features an
overdense bow-shaped head and an underdense tail, which
could be prone to the anti-friction effect.

The AGB outflows deserve extra attention: they are cold ($T
\lesssim 10^3~\K$) and slow (usually subsonic, a few times
$\sim 10~\km~\s^{-1}$), but can still escape from the
stellar potential because of radiation pressure on the
co-moving dust grains \citep[e.g.][]{2017MNRAS.468.4465C,
  2020ApJ...892..110C}. We adopt a simple recipe that the
effective gravitation on the fluids by the primary star is
zero, while the full gravity is still used for the orbital
motion of the binary itself.

\subsection{Fiducial Model Analyses}
\label{sec:agbp-fiducial}

We use the same code unit conversion as in \S\ref
{sec:equimass-fiducial-result} (i.e., $m_c = M_\odot$,
$\ell_c = 1~\au$, $t_c = 1~\yr$), and setup the fiducial
model for AGB-pulsar binary as follows.

Both the primary and the secondary have the same mass as the
Sun, $M_\p = M_\s = M_\odot$, and move at a circular orbit
with separation $D = 5~\au$.  The primary outflow has
mass-loss rate
$\dot{M}_\p = 1.25 \times 10^{-5}~M_\odot~\yr^{-1}$, initial
temperatue $T_{\rm src, p} = 1.4\times 10^3~\K$, and initial
velocity $v_{\rm src, p} = 19~\km~\s^{-1}$ when it leaves
the primary source zone with radius
$r_{\rm src, p} = 0.5~\au$.  The secondary is assumed to be
a windy pulsar, injecting a pulsar wind with initial
temperature $T_{\rm src, s}=10^7~\K$, velocity
$v_{\rm src} = 1.9\times 10^4~\km~\s^{-1}\simeq 0.064 c$,
and $\dot{M}_\s = 3.8\times 10^{-8}~M_\odot~\yr^{-1}$, from
its $R_{\rm src, s} = 0.25~\au$ source zone. As the winds
from both stars can easily escape their potential wells, the
simulation does not need an isothermal condition to sustain
outflows; we use $\gamma = 5/3$ for better capturing the
natures of ionized gas hydrodynamics. These parameters
correspond to (1) a typical AGB wind that disperses
$\sim 50\%$ of its mass during the $\sim 0.5\times 10^5~\yr$
AGB period, and (2) an
$\dot{E} = 4.3\times 10^{36}~\erg~\s^{-1}$ pulsar wind
power, which is on the relatively strong side among the
observed windy pulsars \citep[see e.g.][]
{2011ApJ...726...35A}. Although the spin of pulsars may have
excessively high frequency, it should not affect the binary
evolution in this work due to separation in scales. Assuming
that the pulsar outflow remains rigid co-rotating until the
light cylinder, using the conservation of angular momentum
for the space beyond, we estimate the effect of spin by the
time the pulsar wind reaches the AGB wind,
\begin{equation}
  \label{eq:v-pulsar-spin}
  |\Delta v| \sim \dfrac{S c^2}{2\pi L} \sim
  10~\km~\s^{-1}\times \left( \dfrac{S}{10~{\rm ms}} \right)
   \left( \dfrac{L}{0.1~\au} \right)^{-1}\ ,
\end{equation}
where $S$ is the spin period and $L$ is the typical size of
outflow collision regions. This $|\Delta v|$ is tiny
compared to pulsar winds, allowing us to safely ignore the
spin effects.

The computational box has its side length
$L_{\rm box} = 160~\au$, and the co-rotating Catersian frame
is still applied. Three levels of static mesh refinement
near the secondary and two levels near the primary are
applied to guarantee sufficient resolution: the primary and
secondary source zone radii are resolved by no less than
three cells. Other simulation setups follow
\S\ref{sec:method-setup}. We note that simulations of this
kind are usually difficult because of significant scale
separations in velocities: the speed of binary orbital
motion is slower than the pulsar wind by three orders of
magnitude.  Existing similar simulations are always in 2D
and run for a limited number of periods
\citep[e.g.][]{2012A&A...544A..59B}. Eight NVIDIA RTX 3080
GPUs accelerate these simulations, whose total computing
speed with the \verb|Kratos| code is roughly equivalent
to $\gtrsim 3000$ CPU cores .  It takes $\sim 3.5$ hours
wallclock time for a full orbit in our fiducial model, and
we run the simulation for seven orbital periods
(viz. $t\simeq 7~P$, where $P = 7.91~t_c$).

Figure \ref{fig:agb-pulsar-fiducial} presents the
fundamental hydrodynamic quantities in the orbital
plane. Gas near both objects is in distorted ``tadpole''
shapes. The primary outflow materials dominate the overall
torque, while the secondary outflow maintains the spatial
configurations by providing substantial ram pressure. The
gas velocity exhibits excessive shears near and downstream
of the secondary's terminal shock, leading to significant
instabilities in the secondary's wake. These instabilities
result in large fluctuations of the orbital torque shown in
the upper panel of Figure~\ref{fig:dlns-agbp-fiducial}. The
orbital expansion rate shows a time-averaged positive value
but fluctuates with a standard deviation comparable to the
mean. The lower panel of
Figure~\ref{fig:dlns-agbp-fiducial}, nonetheless, confirms
rapid convergence in terms of spatial coverage, as the
residual of $\Gamma$ still follows the same $\sim r_{\rm
  cut}^{-3}$ scaling law.

If the AGB outflow lasts for a $\sim 0.5\times 10^5~\yr$
period, we can estimate the system will experience an extra
$\sim 6~\%$ orbital expansion using the measured average
expansion rate. During this period, the expansion of the
orbit should also take the mass-loss into account, which is
estimated by the conservation of specific angular momentum,
\begin{equation}
  \label{eq:dlns-mdot}
  \begin{split}
    &\Lambda = [G(M_\p+M_\s)D]^{1/2}\sim {\rm const}\ ,
    \\
    & \dfrac{D_{\rm f}}{D_{\rm i}} \sim
    \dfrac{(M_\p+M_\s)_{\rm i}}{(M_\p+M_\s)_{\rm f}} \ ,
  \end{split}
\end{equation}
where the subscripts ``i'' and ``f'' stand for the initial
and final values, respectively. Consider the fact that the
mass-loss mainly takes place on the primary, after this time
period, the percentage of orbit expansion due to reduced
gravitation (from the $\sim 50~\%$ AGB mass-loss) is
$\sim 1/3$ , and the ``efficiency'' of the trade between
stellar mass and angular momentum is $\sim 20~\%$.

\subsection{Parameter Space and Applications}
\label{sec:agbp-var-models}

\renewcommand{\arraystretch}{1.0}
\begin{deluxetable}{clcc}
  \tablecolumns{4} \tabletypesize{\scriptsize}
  \tablewidth{240pt} \tablecaption{Various models for
    AGB-pulsar binaries. \label{table:agbp-var-model} }
  \tablehead{
    \colhead{Series} & \colhead{Model} &
    \colhead{Description} & \colhead{$\Gamma\ ^*$}
    \vspace*{-0.1in} 
    \\
    \colhead{} &\colhead{} &\colhead{} &
    \colhead{$(10^{-6}~t_c^{-1})$} }
  \startdata
  0 & A0 & Fiducial & $1.13\pm 0.86$ \\
  &&& [ $-1.17\pm 0.65$,\ $2.31\pm 1.01$ ] \\
  \hline
  1 & A1-1 & $\dot{M}_\p=3\times$ Fiducial &
  $2.54\pm 4.19$ \\
  &&& [ $-3.25\pm 1.90$,\ $5.80\pm 3.76$ ] \\
  & A1-2 & $\dot{M}_\p=0.3\times$ Fiducial &
  $-0.11\pm 0.19$ \\
  &&& [ $-0.68\pm 0.28$,\ $0.57\pm 0.16$ ] \\
  \hline
  $2^\dagger$ & A2-1 & $v_{\rm src, s} = 0.1c$ &
  $0.63\pm 0.60$ \\
  &&& [ $-1.68\pm 0.68$,\ $2.31\pm 0.69$ ] \\
  & A2-2 & $v_{\rm src, s} = 0.033c$ &
  $1.46\pm 1.40$ \\
  &&& [ $-0.96\pm 0.59$,\ $2.42\pm 1.33$ ] \\
  \hline
  3$^\ddagger$ & A3 & No AGB $p_{\rm rad}$ &
  $-0.34\pm 0.07$ \\
  &&& [ $-0.40\pm 0.08$,\ $0.06\pm 0.02$ ] \\
  4 & A4 & $M_{\rm s}=2\times M_{\rm p}$ &
  $0.58\pm 0.78$ \\
  &&& [ $-1.10\pm 0.59$,\ $1.68\pm 0.73$ ] \\
  \enddata
  \tablecomments{*: The error shown after each data presents
    the standard deviation for time variation; the data in
    the second row for each model indicates the contribution
    to the overall $(\dot{s}/s)$ from the
    primary (left) and the secondary (right).\\
    $\dagger$: Initial radial velocity of the pulsar wind
    is varied, while the primary wind speed is untouched.\\
    $\ddagger$: Implemented by including the full
    gravitational effects from the primary, in contrast to
    the recipes in the fiducial model that turns off the
    actuall AGB gravitation to emulate an outflow driven by
    radiation pressure $p_{\rm rad}$
    (\S\ref{sec:agbp-fiducial}). }
\end{deluxetable}
 
Similar to \S\ref{sec:agbp-var-models}, we conduct four
extra series of simulations to briefly survey how different
hydrodynamic features impact the orbit evolution, each of
which differs from the fiducial model by only one parameter.
Results of these numerical experiments are presented in
Table~\ref{table:agbp-var-model}. We observe similar trends
as \S\ref{sec:agbp-var-models} in terms of outflow
dependencies, and find that the torque on the primary
generally takes a negative share in the overall balance of
$\Gamma$.

As the lower-right panel in
Figure~\ref{fig:agb-pulsar-fiducial} exhibits, the primary
outflow materials dominate torques on both stars. Higher
primary mass-loss rates generally enhance $\Gamma$ by
increasing torque exerted on the secondary, despite the
offset torque on the primary also rising. Model A1-1, which
follows the same logic, shows a sub-linear growth in the
total torque, as the ratio of primary negative torque
increase is greater than the secondary. Model A1-2, with
$0.3$ times the AGB outflow, yields a total negative torque
that consists of a sharply reduced secondary torque and a
primary negative torque that does not decline
proportionally.

\begin{figure*}
  \centering
  \hspace*{-0.1in} 
  \includegraphics[width=6.5in, keepaspectratio]
  {\figdir/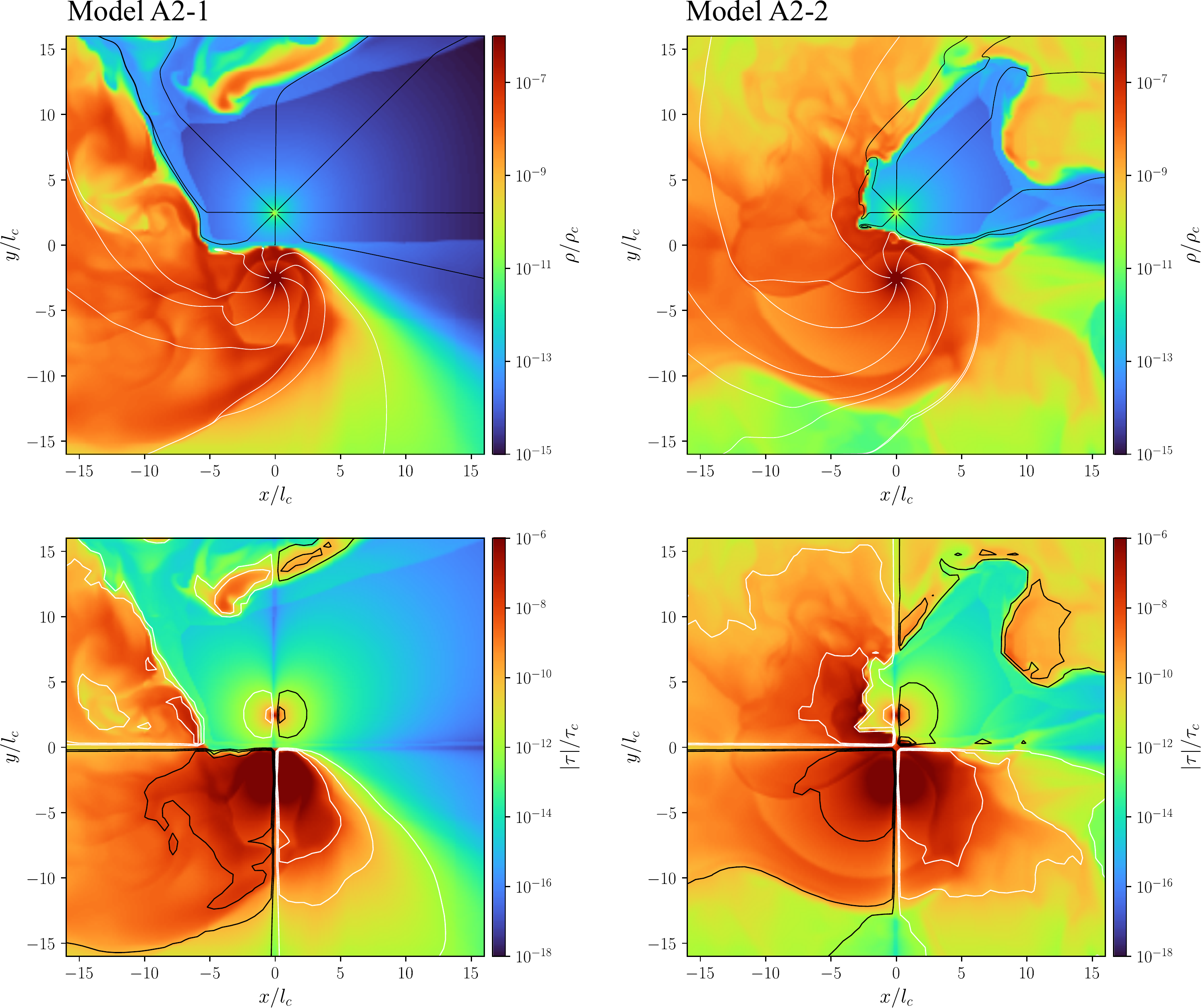}
  \caption{Zoomed-in plots for density (upper row) and
    torque density (lower row) of the $t \simeq 7~P$
    snapshots for Models A2-1 ($v_{\rm src, s} = 0.1~c$,
    left column) and A2-2 ($v_{\rm src, s} = 0.033~c$, right
    column), respectively. Contours in the lower row are
    $\lg|\tau/\tau_c| \in \{ -12, -10, -8 \}$. }
  \label{fig:agb-pulsar-vsec-comp}
\end{figure*}

The same trend is also manifested in Series 2, illustrated
by Figure~\ref {fig:agb-pulsar-vsec-comp}. With higher
pulsar wind speeds, the primary materials are pushed further
from the pulsar but closer to the AGB, enhancing the primary
negative torque and reducing the pulsar's
torque. Considering that the ram pressure is proportional to
the total wind power at a fixed location, this sub-linear
anti-correlation between $\Gamma$ and the pulsar wind speed
indicates that a relatively weaker pulsar wind $\dot{E}$ is
prone to the effect. Specifically, Model A3 reveals the
importance of a sustained, radiation-driven AGB outflow in
this scenario. Once the recipes for radiation pressure
(i.e., zero effective gravity from the primary) is turned
{\it off}, the AGB outflow will have a hard time reaching
the vicinity of the pulsar and remove the base of the
anti-friction effect. Model A4 has doubled primary mass:
under the same mass loss rate, the period to lose $\sim
50~\%$ of its mass is doubled to $\sim
10^5~\yr$. Nevertheless, the eventual amplitude of orbit
expansion is not significantly increased compared to the
fiducial model, even if the integration time is doubled.

\section{Conclusion and Discussions}
\label{sec:discussions}

This paper explores the role of the dynamical anti-friction
effect in binary systems with isotropic outflows from both
stars. We demonstrate that an equal mass binary system could
gain angular momentum through dynamical anti-friction from
the collision between outflows. The resulting expansion of
the orbital radius has amplitude comparable to the expansion
caused by the mass losses. For example, we have shown that
the strong pulsar wind can cause the orbital expansion for
an AGB-pulsar binary.

\subsection{Factors on the Evolution of Orbit Size}
\label{sec:discussion-expansion}

By comparing numerical models, we find that outflows are
crucial to cause the expansion of the orbit. In general, for
each star in a binary, injecting a denser and slower outflow
helps its companion to acquire a greater positive torque in
two ways: first, shaping a denser ``shell'' between the bow
shock and the terminal shock in front of a star for stronger
forward gravity feedback; second, making the direction of
net gravitation feedback force vector closer to tangential
for a greater ``useful'' force component (see also
\S\ref{sec:agbp-var-models}). Nevertheless, the impact of
denser outflows is not always monotonic. Because of the
complicated fluid configurations modulated by orbital
motion, quantitative analyses in, e.g.,
\citet{2020MNRAS.492.2755G} is no longer accurate. Moreover,
when the outflow is dense enough, it will push and
``compress'' the materials behind the gas shell and generate
a denser ``tail'' that offsets the forward force to a
greater extent. Meanwhile, the amount of material reserved
for the outflows could be limited. Since the dynamical
anti-friction effect often increases sub-linearly with
$\dot{M}$, the total effects of $\Delta D$ could even
decrease with increasing $\dot{M}$ in the long
run. Competition between these effects determines the
eventual effects, requiring concrete calculations and
discussions to specify.

Most cases in this paper that have positive $\Gamma$ exhibit
moderate ($\sim 10-30~\%$) rates at which outflowing
binaries ``trade'' their mass for orbital angular
momenta. Because the mass loss will also result in the
expansions of orbits due to weaker gravity, this is not an
impressive efficiency. However, since the sign and amplitude
of $\Gamma$ depend on multiple outflow conditions, the
evolution tracks for two subtlely different binary systems
(although they could be pretty similar in many aspects at
their initial stages) could be convergent or
divergent. These effects have to be taken into account when
one wants to infer the conditions of the progenitors from a
fully-evolved binary. What is more, such evolution should
also be found in the statistics of the evolved and evolving
outflowing binary systems, suggesting more applications of
this mechanism in a broader range of astrophysical
systems. For example, measurements of the orbital properties
for exoplanets are accurate down to $\lesssim 10^{-4}$
nowadays. In the following paper, we shall treat the systems
of exoplanets and their host stars as a particular type of
``binaries'', and further discuss the dynamic effects of the
collision between the young-star stellar wind and the planet
atmospheric outflows.

\begin{figure}[t]
  \centering
  \hspace*{-0.1in} 
  \includegraphics[width=2.8in, keepaspectratio]
  {\figdir/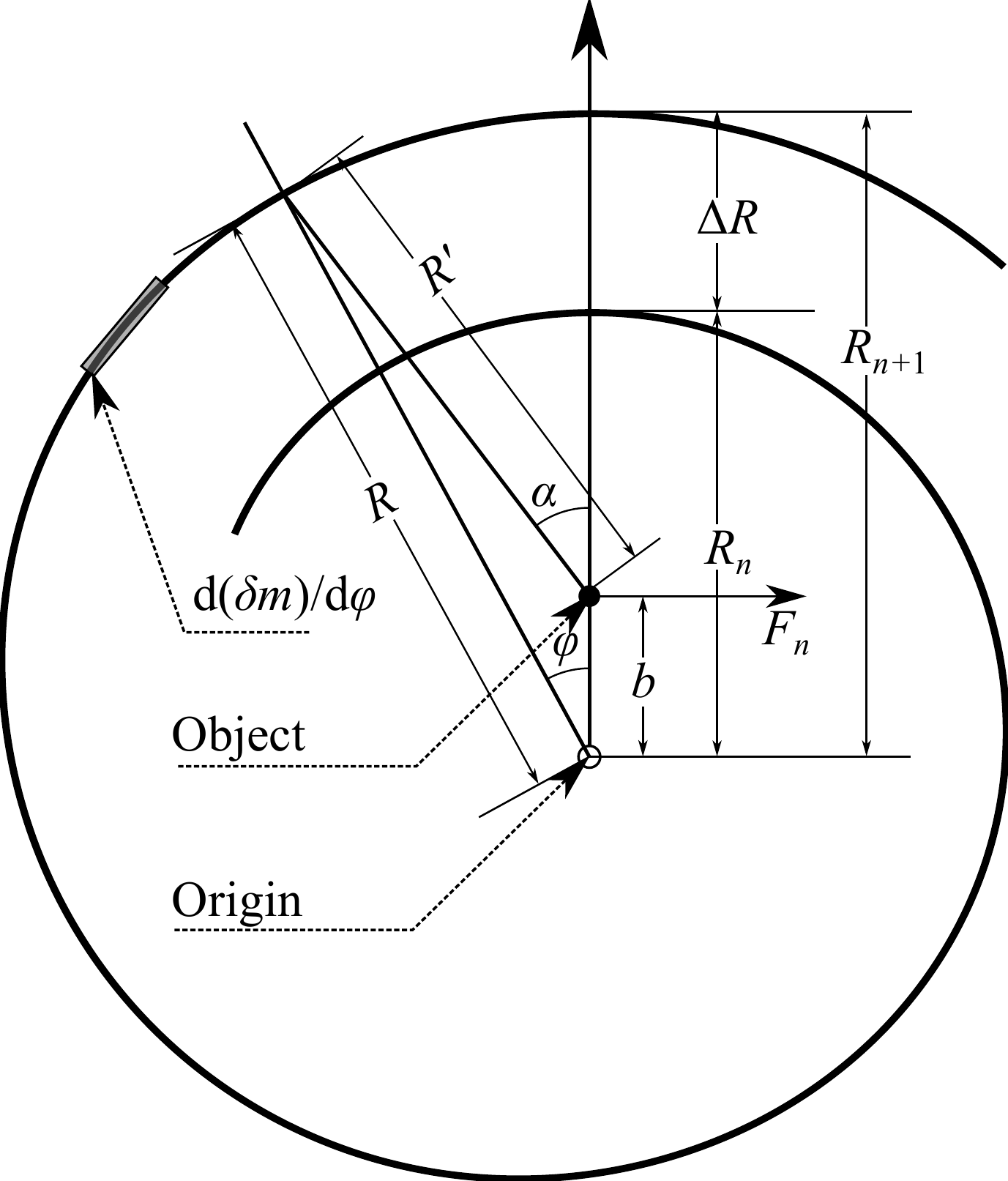}
  \caption{Schematic pattern of spiral-shaped far-field
    perturbation in colliding outflows and its impact to the
    concerned body. Geometric quantities concerned in
    \S\ref{sec:discussions} are all labelled in this figure.
  }
  \label{fig:perturbation-schematic}
\end{figure}

\subsection{Scaling and Convergence of \\
  the Gravitational Effects}
\label{sec:discussion-converge}

One of the common issues that one may raise upon the studies
of dynamical (anti-)friction is the convergence of
gravitational feedback. Because gravitation is a long-range
interaction, correct asymptotes of gravity from materials at
large distances are necessary to yield physically plausible
conclusions; improper cut-offs would impair the reliability
of results. The main concern in \citet{2020MNRAS.494.2327L}
is indeed the divergence of drag force coming from the
gravitational wake due to the simplified assumption of
infinite uniform ambient materials.

There is no such problem for simulations of colliding binary
outflows presented in this paper. The scaling law of
residual torque upon the cutoff radius $r_{\rm cut}$ drops
below the $r_{\rm cut}^{-3}$ envelope, which is a good sign
of convergence (see
e.g. Figures~\ref{fig:dlns-equimass-fiducial} and
\ref{fig:dlns-agbp-fiducial}). This scaling converges much
faster than the $\sim r_{\rm cut}^{-1}$ or even some
divergent scalings in \citet{2020MNRAS.494.2327L}. We now
prove that such $r_{\rm cut}^{-3}$ scaling is not a
coincidence.

To prove the convergence, we focus on the far-field behavior
of the gas. The wake of an outflow-outflow interaction
region moves outwards following the shape of an arithmetic
spiral (see Figure~ \ref{fig:perturbation-schematic}). In
each $(0, 2\pi]$ period of the azimuthal angle $\varphi$
along the spiral, the mass of this wake is simplified into
the linear density distribution along the spiral line
$R(\varphi; n) = R_{n+1} - \varphi \Delta R /(2\pi)$, where
the subscript ``$n$'' marks the $n$th crossing of the
$\varphi = 0$ radial line, and $\Delta R$ is the distance
between adjacent loops. The mass perturbation per unit
azimuthal angle satisfies $\d (\delta m)/\d \varphi \simeq
{\rm const}$, since the evolution of perturbation at large
distances is mostly radial expansion.

Between the $n$-th and the $(n+1)$-th
crossing of the $\varphi = 0$ radial line, the net
tangential force that the interested body feels can be
estimated with simple trigonometries,
\begin{equation}
  \label{eq:net-force-cycle}
  F_n \simeq -G M_*
  \left[\dfrac{\d(\delta m)}{\d \varphi}\right]
    \int_{0}^{2\pi} \dfrac{\sin \alpha\ \d\varphi}{R'^2} \ ,
\end{equation}
where $M_*$ is the object's mass, $b$ is the distance
between the object and the center of mass. $R'$ is the
distance between the line segment and the object, and
$\alpha$ is the direction angle of the line segment seen on
the object's perspective. These geometric quantities are
given by the sine and cosine laws,
\begin{equation}
    \label{eq:spiral-geometry}  
    \begin{split}
      & R'^2 = R^2(\varphi; n) + b^2 - 2 b R(\varphi; n)
      \cos \varphi\ ,
      \\
      & \tan \alpha = \tan \varphi - \dfrac{b}{R
        \cos\varphi}\ .
  \end{split}
\end{equation}
Substituting eq.~\eqref{eq:spiral-geometry} into
eq.~\eqref{eq:net-force-cycle}, and expanding $F_n$ to
the leading-order of $R$, we find $F_n\sim O(R^{-3})$. Since
$R \propto n$ at large $n$, the sum of the series $\{F_n\}$
converges rather quickly with this scaling. This convergence
speed is also applicable in realistic situations where the
density perturbation is three-dimensional: the integral
leading to this scaling is only modified by a bounded
forming factor at the order of unity. From
Figures~\ref{fig:equimass-fiducial} and
\ref{fig:agb-pulsar-fiducial}, we can also confirm this
point clearly from the $\tau$ (``torque density'')
panels. All simulations involved in this work show that the
spatial scaling for the residual gravitational effect upon
the ``cutoff'' radius is indeed $O(r_{\rm cut}^{-3})$. The
outflow-collision scenario resolves the issue of divergence
in the long-range nature of gravity.

\subsection{Future Works}
\label{sec:futures}

In addition to the AGB-pulsar binaries discussed in this
paper, we envisage that dynamical anti-friction may also
cause an orbital expansion in other systems, including the
common envelope evolution and planet migration. This effect
modifies our understanding of binary evolution and may
potentially impact the population modeling of binary compact
object sources for gravitational waves. To simulate the
common envelope evolution and planet systems need special
numerical techniques, which are left for future works.

Our simulations employ a simple approach to assume isotropic
winds are launched from the two stars without specifying the
underlying mechanism. The outflow shuts off the accretion of
materials onto the stars. In reality, the outflows from
either stellar object may be accretion powered and may take
the form of beamed jets that can reduce the effects of
dynamical anti-friction \citep{2020MNRAS.494.2327L}. Future
studies are required to explore the dynamic effects of the
combination of accretion and outflow.

Throughout our simulations, we assume the adiabatic gas
equation of state. However, it is shown that gravitated
outflows with adiabatic index $\gamma > 3/2$ cannot reach
infinity\citep{1958ApJ...128..664P} and heating is generally
a required factor to maintain a smaller effective $\gamma$,
or equivalently a driving mechanism. Such heating could be
rather complicated by radiative transfer, photochemistry,
and thermochemistry. These details could modulate the
thermodynamics dramatically, yielding considerable changes
in the hydrodynamic circumstances that shape the collision
between outflows. Specific to the AGB-pulsar binary models,
both simulation and observation researchers suggest that AGB
outflows are not steady, and significant pulsations could
frequently occur \citep[see e.g.][and references
therein]{2018A&ARv..26....1H, 2021ARA&A..59..337D}. More
thorough and consistent studies in the future should treat
the radiative transfer problem together with the dust grain
formation problem within the AGB winds. We also notice that
the interaction between magnetic fields can re-shape the
stellar winds interactions system, including in the
AGB-pulsar binaries. These details are temporarily ignored
for more explicit analyses and presentations in this work
and also reserve the colorfulness brought by these
subtleties for future works.

\bigskip

L. Wang acknowledges the computation resources provided by
the Kavil Institute of Astronomy and Astrophysics at Peking
University. X. Li is supported by the Natural Sciences and
Engineering Research Council of Canada (NSERC), funding
reference \#CITA 490888-16 and the Jeffrey L. Bishop
Fellowship. Research at Perimeter Institute is supported in
part by the Government of Canada through the Department of
Innovation, Science and Economic Development Canada and the
Province of Ontario through the Ministry of Colleges and
Universities. The authors thank the helpful and inspiring
discussions with our colleagues (alphabetical order of the
last names), Zhuo Chen, Fei Dai.

\bibliography{dyn_friction.bib}
\bibliographystyle{aasjournal}

%
\end{document}